# On the Generalized Bloch Precession Equations

Hans-Jürgen Stöckmann

*Fachbereich Physik der Philipps-Universität Marburg, Renthof 5, D-35032 Marburg*

Dirk Dubbers

*Physikalisches Institut der Universität Heidelberg, Im Neuenheimer Feld 226, D-69120 Heidelberg*

The Bloch equations, which describe spin precession and relaxation in external magnetic fields, can be generalized to include the evolution of polarization tensors of various ranks in arbitrary multipole fields. The derivation of these generalized Bloch equations can be considerably simplified by using a particular bra-ket notation for irreducible tensors.

The *Bloch equations* [1] describe the response of the magnetization $M$ of an ensemble of particles to a classical magnetic field $B$, for the case that the magnetization returns exponentially toward its equilibrium value $M_{z0}$:

$$\dot{M}_x = \gamma(M \times B)_x - M_x / T_2,$$
$$\dot{M}_y = \gamma(M \times B)_y - M_y / T_2, \quad (1)$$
$$\dot{M}_z = \gamma(M \times B)_z - (M_{z0} - M_z) / T_1,$$

with gyromagnetic ratio $\gamma$ and longitudinal and transverse relaxation times $T_1$ and $T_2$.

For negligible relaxation $T_1, T_2 \to \infty$ and written in terms of the vector polarization $P = \langle J \rangle / j$, with angular momentum expectation value $\langle J \rangle$ ($\hbar = 1$), the Bloch equations reduce to the simple precession equation

$$\dot{P} = P \times \omega_L, \quad (2)$$

with the Larmor angular frequency vector $\omega_L = \gamma B$. This equation holds for arbitrary angular momentum quantum number $j \geq \tfrac{1}{2}$, and is all one needs for a complete quantum description of this simple system. Equation (2) is called semiclassical only because the external field $B$ is not quantized, i.e., there is no back action of the spins on the magnetic field.

While Eqs. (1) and (2) were originally derived in the context of magnetic resonance, it was soon recognized that they describe the evolution of any system with "effective" or "pseudospin" $J$, the best known examples being the maser [2], or the two atomic states involved in an optical transition [3], which obey the "optical Bloch equations", both systems having effective spin-½. Even multiple-quantum transitions are covered by these equations.

The Bloch equations can be generalized to include the evolution of the higher multipole moments of spin polarization in the presence of arbitrary multipole fields. Starting point is the Liouville equation

$$\dot{\rho} = i[\rho, H] \quad (3)$$

for the density matrix $\rho$. For a given angular momentum state $j$, its rank is $2j+1$, with matrix elements $\langle jm | \rho | jm' \rangle$.

The Liouville equation (3) may be rewritten in terms of a *generalized spin precession equation*

$$\dot{\rho}_L = i \sum_{L_1 L_2} c_j(L_1, L_2, L) [\Omega_{L_1} \times \rho_{L_2}]_L, \quad (4)$$

with the statistical tensors $\rho_L$, also called state multipoles, with $2L+1$ polarization coefficients $\rho_{LM}$, $M = -L, ..., L$, and with the interaction tensors $\Omega_{L_1}$, with $2L_1 + 1$ frequencies $\Omega_{L_1 M_1}$. These tensors will be defined below, as will be their products and the coefficients $c_j$. For the special case that there are only interactions with uniform magnetic fields, Eqs. (4) reduce to Eq. (2). Eqs. (4) had been derived by Fano [4] in an important paper, which regrettably never found the recognition it deserves. A possible explanation for this may lay in the fact that it uses an unfamiliar notation, and that the derivation is technical and tends to obscure the essential underlying ideas.

The purpose of the present letter is to show that the derivation of Eq. (4) from Eq. (3) can be considerably simplified and becomes nearly trivial by using a special bra-ket notation for the irreducible tensors. With this notation Eq. (4) will be obtained by a number of simple steps involving essentially the Wigner-Eckart theorem and nothing else.

The same procedure then is applied to derive the equivalent of Eqs. (1), namely the *generalized Bloch equations*, which, under conditions to be discussed later on, read

$$\dot{\rho}_{LM} = i \sum_{L_1 L_2} c_j(L_1, L_2, L) [\Omega_{L_1} \times \rho_{L_2}]_{LM} - \frac{\rho_{LM}}{\tau_{LM}}, \quad (5)$$

with separate relaxation times $\tau_{LM}$ for each polarization component $\rho_{LM}$.

The irreducible elements $\rho_{LM}$ of the density operator are obtained from its cartesian components by

$$\rho_{LM} = \sum_{mm'} \sqrt{\frac{2L+1}{2j+1}} \langle jmLM | jm' \rangle \langle jm | \rho | jm' \rangle, \qquad (6)$$

where $\langle jmLM | jm' \rangle$ is a Clebsch-Gordon coefficient. The rank of the $\rho_L$ is limited to $L \leq 2j$, as follows from the triangular rule for the Clebsch-Gordon coefficients. The rank-one polarization vector $\rho_1$ is often called *orientation* or simply *polarization*, and the rank-two tensor $\rho_2$ *alignment*.

In Eq. (4), the tensor products $[\Omega_{L_1} \times \rho_{L_2}]_L$, with $|L_1 - L_2| \leq L \leq L_1 + L_2$, are the generalization of the vector product $\mathbf{P} \times \boldsymbol{\omega}_L$ in Eq. (2). The elements of the tensor product are defined by the bilinear form

$$[\Omega_{L_1} \times \rho_{L_2}]_{LM} = \sum_{M_1 M_2} \langle L_1 M_1 L_2 M_2 / LM \rangle \, \Omega_{L_1 M_1} \rho_{L_2 M_2}. \qquad (7)$$

A main use of the statistical tensor elements $\rho_{LM}$ is in the angular distribution or correlation of particles emitted in nuclear, atomic, or molecular reactions, see for instance ch. 19 of [5]. These distributions can be written in terms of the elements of the rotation matrices, for instance, for the case of a rotationally symmetric configuration, as

$$W(\theta) = \sum_L r_L \, \rho_{L0} \, P_L(\cos\theta), \qquad (8)$$

with Legendre polynomials $P_L$ and some coefficients $r_L$.

Equations (3) and (4) are mathematically equivalent, but the spin precession equation has a number of obvious advantages as compared to the original Liouville equation:

(i) The $\rho_{LM}$ allow a direct interpretation in terms of polarization, alignment etc. Very often just one tensor component is prepared in the production of the ensemble, all other initial components being zero. The spin precession equation is ideally suited to cope with this situation.

(ii) The tensor products allow an easy visualization of what is going on. From the selection rules of the Clebsch-Gordon coefficients and the prefactors $c_j$ (see Eqs. (28) and (29) below) it is obvious, e.g., that a magnetic interaction can only change the $M$ state of a tensor component, but never its rank $L$, whereas a quadrupole interaction always changes the rank by $\pm 1$.

(iii) If there are symmetries, e. g., rotation symmetries about the axis of an external magnetic field, the generalized spin precession equation systems for $(2j+1)^2$ variables will decompose into smaller uncoupled differential equations.

To derive the generalized precession equations we first recall the definition of an irreducible tensor operator via its commutation relations with the elements $L_z$ and $L_\pm$ of the orbital angular momentum operator

$$\begin{aligned}[]
[L_z, T_{LM}] &= M \, T_{LM}, \\
[L_\pm, T_{LM}] &= \sqrt{L(L+1) - M(M \pm 1)} \, T_{L,M\pm 1}.
\end{aligned} \qquad (9)$$

The close similarity between this definition of the tensor operators $T_{LM}$ and the well known relations for the eigenstates $|lm\rangle$,

$$\begin{aligned}
L_z | lm \rangle &= m | lm \rangle, \\
L_\pm | lm \rangle &= \sqrt{l(l+1) - m(m \pm 1)} \, | l, m \pm 1 \rangle,
\end{aligned} \qquad (10)$$

can be made even more suggestive by introducing a special notation for the tensor operators, see also [6]. We define tensor bras and kets (using round brackets)

$$|LM) = T_{LM}, \quad (LM| = T_{LM}^\dagger. \qquad (11)$$

At the same time, we assume that an operator $A$ acts on the tensor bras and kets as

$$A|LM) = [A, T_{LM}], \quad (LM|A = [T_{LM}^\dagger, A]. \qquad (12)$$

With these definitions, Eqs. (9) and (10) become identical even with respect to notation. One only has to replace the bras and kets in Eqs. (10) by the corresponding tensor bras and kets, i.e., Eqs. (12) with $A = L_z, L_\pm$, to recover Eqs. (9) for the tensor operators.

We then define the tensor matrix element of an operator $A$ as

$$\begin{aligned}
(L'M' | A | LM) &= \mathrm{Tr}\{T_{L'M'}^\dagger [A, T_{LM}]\} \\
&= \mathrm{Tr}\{[T_{L'M'}^\dagger, A] T_{LM}\}.
\end{aligned} \qquad (13)$$

The last identity holds because of the commutativity property of the trace. It is thus irrelevant whether operator $A$ acts on the bra or on the ket.

The matrix elements of the product $AB$ of two operators $A$ and $B$ follow from the completeness relation for the irreducible tensor operators as

$$\begin{aligned}
(L_1 M_1 &| AB | L_2 M_2) \\
&= \sum_{LM} (L_1 M_1 | A | LM) \, (LM | B | L_2 M_2),
\end{aligned} \qquad (14)$$

as can be seen by writing down explicitly all matrix elements from Eq. (13), where the application of the product $AB$ on $|LM)$ does not mean $[AB, T_{LM}]$, but has to be interpreted as

$$AB|LM) = [A, [B, T_{LM}]]. \qquad (15)$$

The $T_{LM}$ are orthogonal with respect to the trace operation,

$$\mathrm{Tr}(T_{LM}^\dagger T_{L'M'}) = \mathrm{Tr} |\mathrm{T}_L|^2 \delta_{LL'} \delta_{MM'}, \qquad (16)$$

where $\text{Tr}|T_L|^2$ is independent of $M$ as is indicated by the notation. The proof follows exactly the same line as the corresponding proof for the orthogonality of the eigenfunctions $|lm\rangle$ which can be found in any textbook. Throughout this paper we assume that the $T_{LM}$ are normalized, i.e., $\text{Tr}|T_L|^2 = 1$. For $L = 0$ to 2 the normalized tensors are given by

$$T_{00} = a_0 I, \quad T_{10} = a_1 J_z, \quad T_{1\pm 1} = \mp \sqrt{\tfrac{1}{2}}\, a_1 J_\pm,$$
$$T_{20} = a_2(3J_z^2 - J^2), \quad T_{2\pm 1} = \mp \tfrac{1}{2}\sqrt{6}\, a_2 (J_\pm J_z + J_z J_\pm), \quad (17)$$
$$T_{2\pm 2} = \tfrac{1}{2}\sqrt{6}\, a_2 J_\pm^2,$$

with identity $I$ and

$$a_0 = \sqrt{1/(2j+1)}, \quad a_1 = \sqrt{3/j(j+1)(2j+1)},$$
$$a_2 = \sqrt{20/(2j-1)2j(2j+1)(2j+2)(2j+3)}. \quad (18)$$

The matrix elements of the $T_{LM}$ in the $|jm\rangle$ basis are easily calculated by means of the Wigner-Eckart theorem,

$$\langle jm'|T_{LM}|jm\rangle = \langle j||T_L||j\rangle \langle jmLM|jm'\rangle / \sqrt{2j+1}, \quad (19)$$

where the reduced matrix element shows up to be $\langle j||T_L||j\rangle = \sqrt{2L+1}$, as an immediate consequence of the orthogonality relation for the Clebsch-Gordon coefficients.

We are now prepared to derive the generalized spin-precession equation (4) from the Liouville equation (3). To this end we expand the density matrix in terms of normalized irreducible tensor operators,

$$\rho = \sum_{LM} \rho_{LM} T_{LM}^\dagger. \quad (20)$$

Inversion of Eq. (20), using the orthogonality of the tensor operators, yields

$$\rho_{LM} = \text{Tr}(\rho T_{LM}) = \langle T_{LM}\rangle. \quad (21)$$

The latter equation is identical with the previous definition (6) of the state multipoles. The $\rho_{LM}$ are thus nothing but the expectation values of the normalized irreducible tensors.

We then take the time derivative on both sides of Eq. (21), and on the right hand side we insert Eqs. (3), and subsequently Eq. (20), which leads to

$$\dot\rho_{LM} = \text{Tr}(\dot\rho\, T_{LM}) = i\,\text{Tr}([\rho,H]T_{LM})$$
$$= i \sum_{L_2 M_2} \rho_{L_2 M_2} \text{Tr}([T_{L_2 M_2}^\dagger, H] T_{LM}). \quad (22)$$

With the bracket notation (13) for the tensor matrix elements, we can write Eq. (22) simply as

$$\dot\rho_{LM} = i \sum_{L_2 M_2} \rho_{L_2 M_2} (L_2 M_2 | H | LM). \quad (23)$$

We then expand also the Hamiltonian $H$ in terms of normalized irreducible tensor operators,

$$H = \sum_{L_1 M_1} \Omega_{L_1 M_1} T_{L_1 M_1}^\dagger. \quad (24)$$

The interaction frequencies therein are, for the mainly occurring magnetic dipole and electric quadrupole interactions,

$$\Omega_{10} = \gamma B_z/a_1, \quad \Omega_{1,\pm 1} = \mp \gamma(B_x \pm iB_y)/\sqrt{2}\,a_1,$$
$$\Omega_{20} = (\omega_Q / a_2)(2\varphi_{zz} - \varphi_{xx} - \varphi_{yy})/3\varphi_{zz},$$
$$\Omega_{2\pm 1} = \mp(\omega_Q / a_2)(\varphi_{zx} \pm i\varphi_{zy})/\sqrt{6}\,\varphi_{zz}, \quad (25)$$
$$\Omega_{2\pm 2} = (\omega_Q / a_2)(\varphi_{xx} - \varphi_{yy} \pm 2i\varphi_{xy})/\sqrt{6}\,\varphi_{zz},$$

with the quadrupole interaction frequency $\omega_Q = e\varphi_{zz}Q/4j(2j-1)$, and with the second spatial derivatives $\varphi_{xy} = \partial^2\varphi/\partial x \partial y$, etc., of the external electric potential $\varphi$, which constitute the elements of the electric field gradient tensor.

Inserting Eq. (24) into Eq. (23) gives

$$\dot\rho_{LM} = i \sum_{L_1 M_1 L_2 M_2} \Omega_{L_1 M_1} \rho_{L_2 M_2} (L_2 M_2 | T_{L_1 M_1}^\dagger | LM). \quad (26)$$

Next we have to evaluate the tensor matrix element $(L_2 M_2 | T_{L_1 M_1}^\dagger | LM) = (LM | T_{L_1 M_1} | L_2 M_2)^*$ with round brackets. Because of the one-to-one correspondence of (9) and (10), each proof for the tensor functions is automatically true for the tensor operators, too, and vice versa. In particular, we can directly apply the Wigner-Eckart theorem to calculate the matrix elements with tensor bra-kets

$$(LM | T_{L_1 M_1} | L_2 M_2)$$
$$= (L || T_{L_1} || L_2)\langle L_2 M_2 L_1 M_1 | LM\rangle / \sqrt{2L+1}. \quad (27)$$

Entering Eq. (26) with expression (27) for the tensor matrix elements, one immediately arrives at the generalized spin precession equation (4) with

$$c_j(L_1 L_2 L) = -(L || T_{L_1} || L_2)/\sqrt{2L+1}. \quad (28)$$

By a consequent use of the normalized irreducible tensors and the round-bracket tensor matrix elements, the derivation of the generalized spin precession equation (4) has been reduced to a small number of elementary steps. Only the Wigner-Eckardt theorem has been used in the derivation. Since this is a universal theorem holding in all symmetry groups, the generalized spin precession equation, too, is applicable in all groups. The individual group structures only enter in the calculation of the reduced matrix element in Eq. (28). It is only here where some computational effort is needed [4], see also Appendix A.3 of [7]. If this is done, one obtains

$$(L//\mathrm{T}_{L_1}//L_2) = (-1)^{2j+L} [(-1)^{L_1+L_2-L} - 1]$$
$$\times \sqrt{(2L_1+1)(2L_2+1)(2L+1)} \begin{Bmatrix} L_1 & L_2 & L \\ j & j & j \end{Bmatrix}. \quad (29)$$

In the literature, statistical tensors are generally employed without using Fano's compact and suggestive tensor product expression (4). For the purely magnetic case, magnetic resonance line shapes of the $\rho_{LM}$ were treated in [8], including the line shapes of atomic double-resonance signals as known from [9]. The Hanle effect as based on atomic alignment was studied in [10]. For further experiments on the reorientation of atomic state multipoles, see [11, 12], and references therein.

Let us now turn to the discussion of the relaxation term in Eq. (5). The relaxation of the $\rho_{LM}$ was first treated in [13], and independently in [14]. For more recent uses of the statistical tensors in magnetic resonance and relaxation, see [15-17], and the review [18]. In relaxation studies, the tensor bra-ket notation exhibits its full power. It is the usual situation in every experiment that there is a system Hamiltonian $H_S$, which is under control of the experimentalist, the environment described by a bath Hamiltonian $H_B$, which usually cannot be controlled, and a Hamiltonian $H_{SB}$ coupling the system to the environment. Removing the bath Hamiltonian $H_B$ by a standard transformation one obtains an effective Hamiltonian $H = H_S + H_{SB}(t)$, where the coupling Hamiltonian $H_{SB}(t) = \exp(iH_B t) H_{SB}(0) \exp(-iH_B t)$ now has turned into a time-dependent interaction. After another transformation

$$\hat{\rho} = e^{iH_S t} \rho \, e^{-iH_S t} \quad (30)$$

we arrive at the Liouville equation in the interaction representation,

$$\dot{\hat{\rho}}_{LM} = i \sum_{L'M'} \hat{\rho}_{L'M'} (L'M'|\hat{H}_1(t)|LM) \quad (31)$$

obtained from Eq. (23) be replacing $\rho$ by $\hat{\rho}$ and $H$ by

$$\hat{H}_1(t) = e^{iH_S t} H_{SB}(t) \, e^{-iH_S t}. \quad (32)$$

It is beyond the scope of this letter to go into the details of relaxation theory. Instead we just follow the standard approach and treat relaxation in second order perturbation theory. To this end we integrate both sides of Eq. (31) over $t$,

$$\hat{\rho}_{L'M'}(t) = \hat{\rho}_{L'M'}(0)$$
$$+ i \sum_{L''M''} \int_0^t d\tau \, \hat{\rho}_{L''M''}(\tau) (L''M''|\hat{H}_1(\tau)|L'M'), \quad (33)$$

where in addition we relabelled the indices, and substitute this expression for $\hat{\rho}_{L'M'}(t)$ on the right hand side of Eq. (31),

$$\dot{\hat{\rho}}_{LM}(t) = i \sum_{L'M'} \hat{\rho}_{L'M'}(0) \, (L'M'|\hat{H}_1(t)|LM)$$
$$- \sum_{L'M'L''M''} \int_0^t d\tau \, \hat{\rho}_{L''M''}(\tau)$$
$$\times (L''M''|\hat{H}_1(\tau)|L'M') \, (L'M'|\hat{H}_1(t)|LM)$$
$$= i \sum_{L'M'} \hat{\rho}_{L'M'}(0) \, (L'M'|\hat{H}_1(t)|LM)$$
$$- \sum_{L'M'} \int_0^t d\tau \, \hat{\rho}_{L'M'}(t-\tau)$$
$$\times (L'M'|\hat{H}_1(t-\tau)\hat{H}_1(t)|LM), \quad (34)$$

where in the second step the completeness of the irreducible tensor operators has been used. In addition we changed the integration variable from $\tau$ to $t - \tau$. To avoid a possible misinterpretation of the equation it is reminded of the operator multiplication convention (15).

Equation (34) is still exact, but to proceed further we have to apply approximations: (i) First we assume that $\hat{H}_1(t)$ varies rapidly with time as compared to $\hat{\rho}_{L'M'}(t)$. Then we may average the right hand side over these rapid fluctuations by replacing $\hat{H}_1(t)$ and $\hat{H}_1(t-\tau)\hat{H}_1(t)$ by their time averages $\overline{\hat{H}_1(t)}$ and $\overline{\hat{H}_1(t-\tau)\hat{H}_1(t)}$, respectively. (ii) Without loss of generality we may assume that $\overline{\hat{H}_1(t)}$ disappears, if needed after a proper redefinition of the system Hamiltonian $H_S$. (iii) Next, the term $\overline{\hat{H}_1(t-\tau)\hat{H}_1(t)}$ typically will decay on a time scale $\tau_c$, the correlation time for the fluctuations of the bath variables. If $\tau_c$ is short compared to the time scale where the changes of $\hat{\rho}_{LM}(t)$ take place, we may replace $t - \tau$ in the argument of $\hat{\rho}_{L'M'}$ on the right hand side by $t$, and extend the upper limit of the integration to $\infty$. (iv) Finally, for the sake of simplicity let us restrict the sum over $(L'M')$ to just one term $(LM)$. We then obtain

$$\dot{\hat{\rho}}_{LM} = -\hat{\rho}_{LM}/\tau_{LM}, \quad (35)$$

where

$$\frac{1}{\tau_{LM}} = \int_0^\infty d\tau \, (LM|\overline{\hat{H}_1(t-\tau)\hat{H}_1(t)}|LM). \quad (36)$$

Going back to the laboratory system by inverting transformation (30) we obtain the generalized spin precession equation (33) including the relaxation term. We see that on the level of the applied approximations, each tensor component decays exponentially with its own decay constant, but this is no longer true for more evolved relaxation theories, in particular if fluctuating quadrupole interactions are involved [14]. For tensor rank $L = 1$ we recover Bloch's equation, where we can identify $\tau_{10}$ and $\tau_{1\pm1}$ with $T_1$ and $T_2$, respectively. We do not recover, however, a decay towards a thermal equilibrium in this way, since all tensor components at the very end decay to zero. A better approach repairing this deficiency would take into account the Boltzmann polarization of the bath, which is not difficult to do, but not of relevance in the present context. To get explicit expressions for the relaxation rates from Eq. (36) for various situations found in experiments still some effort is needed [19].

In conclusion, using a special bra-ket notation, the derivation of the generalized Bloch precession equations from the Liouville equation has been reduced to three to four nearly trivial steps, both for the precession and the relaxation term. The key ingredients were the introduction of the operator tensor matrix elements, Eq. (14), and the reinterpretation of the statistical tensors $\rho_{LM}$ in terms of expectation values of the normalized irreducible tensor, Eq. (21). Since nothing but the Wigner-Eckardt theorem (27) was used in the derivation, Eq. (4) is valid for arbitrary groups. Only in the calculation of reduced matrix elements, Eq. (29), or the explicit evaluation of the relaxation rates from Eq. (36), the individual group structures enter. Though it is only a special notation we used in this letter, we consider the advantages of the present approach as so convincing that the technique still should be of use for all people working in the field, in spite of the fact that nearly half a century has passed since Fano's original work.

[1] Bloch, F., Phys. Rev. **70**, 460 (1946).
[2] Feynman, R.P., Vernon, F.L. Jr., Hellwarth, R.W., J. Appl. Phys. **28**, 49 (1957).
[3] Allen, L., and Eberly, J.H., *Optical Resonance and Two-Level Atoms* (Dover Publications, New York 1975).
[4] Fano, U., Phys. Rev. **133**, B828 (1964).
[5] Fano, U., and Racah, G., *Irreducible Tensorial Sets* (Academic Press, New York 1959).
[6] Sanctuary, B.C., J. Magn. Res. **61**,116 (1985).
[7] Dubbers, D., and Stöckmann, H.-J., *Quantum Physics: The Bottom-Up Approach* (Springer, Heidelberg 2013).
[8] Matthias, E., *et al.*, Phys. Rev. A **4**, 1626 (1971).
[9] Brossel, J., and Bitter, F., Phys. Rev. **86**, 308 (1952).
[10] Breschi, E., and Weis, A., Phys. Rev. A **86**, 053427 (2012).
[11] Blum, K., *Density Matrix Theory and Applications* (Plenum Press, New York 1996).
[12] Bain, A.D., and Berno, B., Progr. Nucl. Magn. Spectr. **59**, 223 (2011).
[13] Gabriel, H., Phys. Rev. **181**, 506 (1969).
[14] Happer, W., Phys. Rev. B **1**, 2203 (1970).
[15] Kruk, D., J. Chem. Phys. **135**, 224511 (2011).
[16] Ramachandran, R., and Griffin, R.G., Molec. Phys. **122**, 164502 (2005).
[17] Gustavsson, S., and Halle, B., Molec. Phys. **80**, 549 (1993).
[18] Nielsen, R.D., and Robinson, B.H., Concepts Magn. Res. **28A**, 270 (2006).
[19] Abragam, A., *The Principles of Nuclear Magnetism*, (Oxford University Press, Oxford 1961).